\documentclass[english,american]{elsarticle}
\usepackage{mathptmx}
\usepackage{helvet}
\usepackage{courier}
\usepackage[T1]{fontenc}
\usepackage[latin9]{inputenc}
\usepackage[a4paper]{geometry}
\usepackage{amsmath}
\usepackage{graphicx}

\makeatletter
\usepackage[scanall]{psfrag}

\let\jmath=\undefined
\DeclareSymbolFont{cmletters}{OML}{cmm}{m}{it}
\DeclareMathSymbol{\jmath}{\mathord}{cmletters}{"7C}

\def\citet#1{\cite{#1}}
\date{}

\makeatother

\usepackage{babel}
\begin{document}

\title{Extracting the contribution of conduction in permittivity measurements
using the Kramers-Kronig relations}

\author{\textbf{\normalsize{C. Margo$^{1}$,E. Géron$^{1}$,T.Ditchi , S.Holé$^{2}$
and J.Lucas$^{1*}$}}{\normalsize{}}\\
{\normalsize{Laboratoire de Physique et d\textquoteright{}Etude des
Matériaux, CNRS UMR8213, France}}\\
{\normalsize{$^{*}$ Corresponding author : jerome.lucas@espci.fr,
+33 1 40 79 45 33}}\\
{\normalsize{$^{1}$ESPCI-ParisTech,}}\\
{\normalsize{$^{2}$UPMC Sorbonne Universités}}}
\begin{abstract}
The measurement of the material permittivity is often performed via
an impedance measurement. In this case the measured permittivity includes
the conduction contribution. Most of the time, the impedance-meters
performing such measurement do not perform static measurement even
if they can go to low frequencies. When the dipolar relaxation of
the material leads to low frequency relaxation, separating the relaxation
phenomenon from the conduction phenomenon can be difficult, especially
when multiple relaxation phenomena superimpose. In this work we use
the Kramers-Kronig relations to perform that separation by considering
the frequency response aspect of the permittivity. The practical aspect
of the method is presented and demonstrated on real measurements.
\end{abstract}
\maketitle

\section{Introduction}

Measuring the conductivity of poorly conducting material such as cement
is a difficult task. In this paper, we consider measurements performed
using an impedance meter. The measurement of theses devices are performed
by exiting a dielectric sample with potential at given frequencies
and measuring the resulting current. A dielectric constant measurement
can then be elaborated from the probe geometry and sample thickness~\citet{Agilent2013}.
Whatever the nature of the sample, the measured current is the sum
of the conduction current and of the displacement current~\citet{Kittel1976}.
In a resistive and linear material, the conduction current is preponderant
at low frequencies when the influence of the displacement current
is effective around the dipolar relaxation frequencies of the material~\citet{Jackson1967,Ramo1984}.
Furthermore, when the relaxation frequencies are low as in the case
of cement, both effect are superimposed making their separation difficult.
In this paper, after a short review of the Debye model that we use
as a reference, we show that the Kramer-Kronig relations~\citet{BOHREN2010}
allow to split those effects with no assumptions on the relaxation
model even when the relaxation frequencies are very low. Finally,
as an illustration, the method is applied to some typical measurements
obtained with a cement sample using the technique developed in \citet{Lucas2012}.

\section{Theoretical Background}

\subsection{Complex permittivity of dielectrics\label{sub:ComplexPermittivity}}

Permittivity $\epsilon(\omega)$ corresponds to the response of a
material to the electrical field. Dipoles get oriented more or less
rapidly depending on their interactions with the lattice. As the displacement
field $\vec{D}=\epsilon\vec{E}$ is the response of the material,
it is necessarily out of phase with the electrical field. Permittivity
$\epsilon(\omega)$ must be then a complex function to describe the
phenomenon correctly. 

For an isotropic homogeneous material whose response to the electrical
field is linear and containing only one kind of dipoles, the Debye
model applies and the permittivity can be written for one relaxation
frequency as
\begin{equation}
\epsilon_{D}(\omega)=\epsilon_{\infty}+\frac{\epsilon_{s}-\epsilon_{\infty}}{1+\jmath\omega/\omega_{\mathrm{rel}}}\label{eq:Debye}
\end{equation}
where $\omega$ is the circular frequency, $\omega_{\mathrm{rel}}$
the dipolar relaxation circular frequency, $\epsilon_{s}$ is the
static dielectric constant, $\epsilon_{\infty}$ is the optic region
permittivity and $\jmath=\sqrt{-1}$. Notice that there may be more
than one relaxation frequency if there are other kinds of dipoles
involved in the process. The dipolar relaxation phenomenon appears
at low frequency for large molecules. It appears at higher frequencies
for smaller molecules or fractions of large molecules and even at
higher frequencies at the level of atoms \citet{Jackson1967}.

\subsection{Impedance measurement of permittivity}

\begin{figure}[h]
\centering{}\fontsize{8pt}{10pt}\selectfont
\psfrag{Real}{Real part of $\epsilon (\omega)/\epsilon _0$}
\psfrag{Imag}{Imagenry part of $\epsilon (\omega)/\epsilon _0$}
\psfrag{F/m}[cc][cc]{}
\psfrag{-1}{$0.1$}
\psfrag{0}{$1$}
\psfrag{1}{$10$}
\psfrag{2}{$100$}
\psfrag{3}{$10^{3}$}
\psfrag{Ghz}{GHz}
\psfrag{4.35}{}
\psfrag{4.3}[l][l]{4.3}
\psfrag{4.25}{}
\psfrag{4.2}{4.2}
\psfrag{4.15}{}
\psfrag{4.1}{4.1}
\psfrag{4.05}{}
\psfrag{0.45}{}\psfrag{KK from Debye}{Calculated with K-K}
\psfrag{0.4}{0.4}
\psfrag{0.35}{}
\psfrag{0.3}{0.3}
\psfrag{0.25}{}
\psfrag{0.2}{0.2}
\psfrag{0.15}{}
\psfrag{0.1}{0.1}
\psfrag{0.05}{}
\psfrag{0.0}{0.0}
\psfrag{Debye}{Polarization and conduction}
\psfrag{DebyeNoDC}{Polarization only}
\psfrag{KK from Debye}{Calculated with K-K}
\includegraphics[width=8cm]{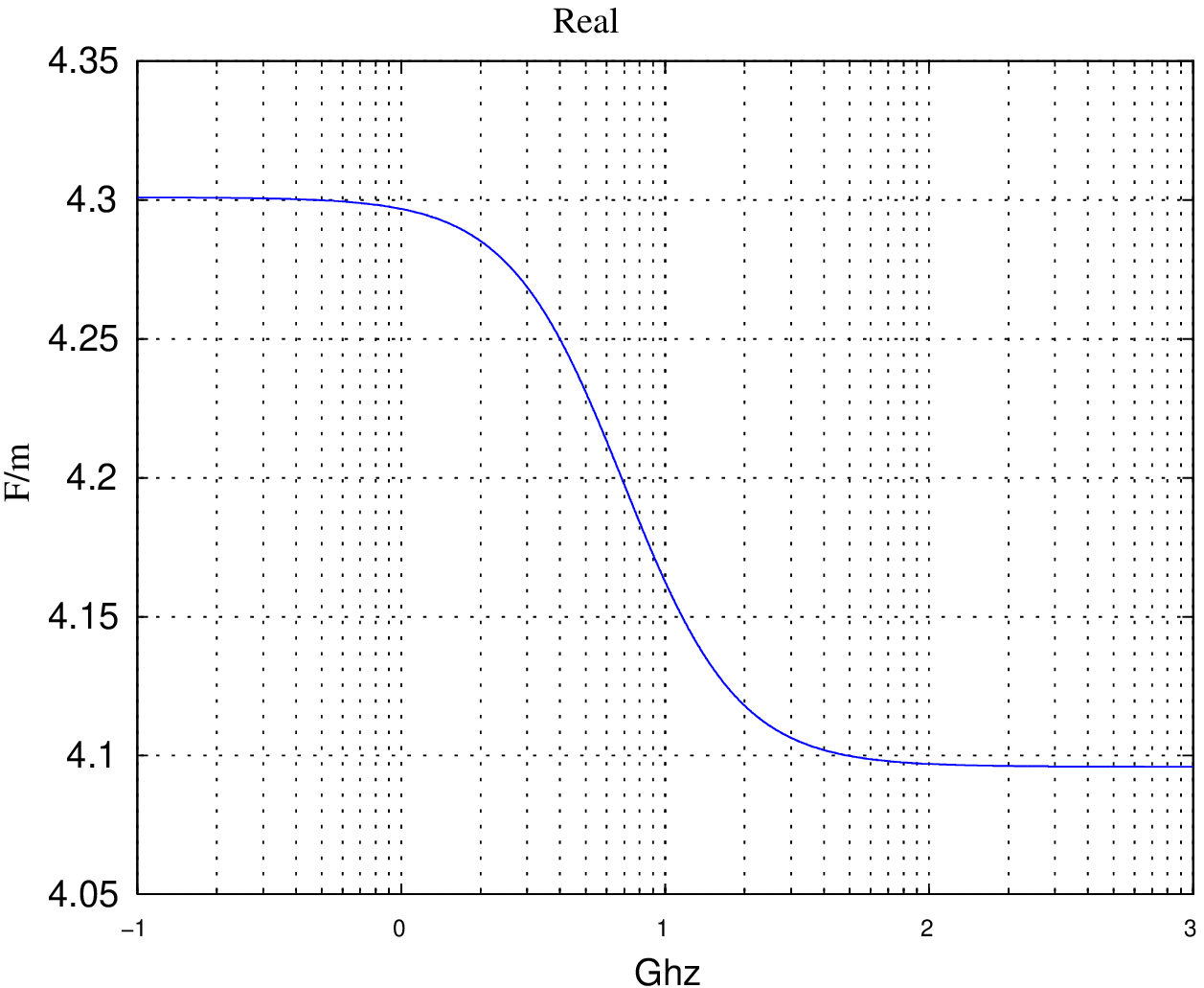}~~\includegraphics[width=8cm]{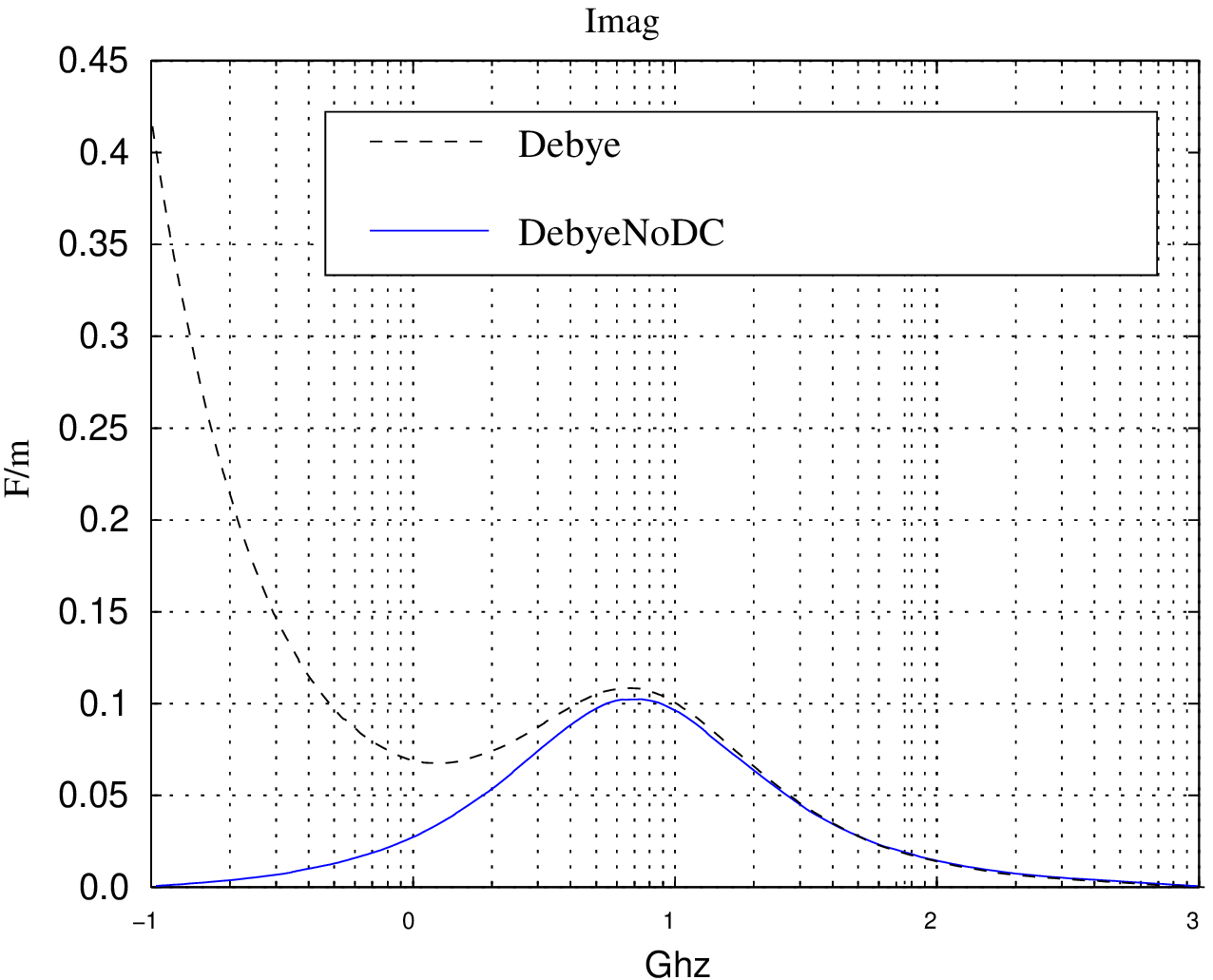}\caption{Permittivity of FR4 substrate accordingly to the Debye Model.\label{fig:DebyeModelFR4}}
\end{figure}

The displacement field $\vec{D}$ is difficult to measure directly.
The permittivity is then often measured through a current measurement.
The studied material is placed in a sample holder of known geometry,
and the build capacitor is polarized under $V(\omega)$. By measuring
the current $I(\omega)$, one can obtain the capacitance from $\jmath\omega C(\omega)=I(\omega)/V(\omega)$
and then deduces $\epsilon(\omega)$ knowing the capacitor geometry.
Unfortunately, the measured current consists in a contribution of
the displacement current density $\partial\vec{D}/\partial t$ and
a contribution of the conduction current density $\sigma\vec{E}$
due to the free charges present in the material event if they are
few. Consequently the measured permittivity can be written as:
\begin{equation}
\epsilon_{M}(\omega)=\epsilon_{\infty}+\Big[\epsilon_{P}^{'}(\omega)+\jmath\epsilon_{P}^{''}(\omega)\Big]-\jmath\frac{\sigma}{\omega}\label{eq:FullDebye}
\end{equation}
where $\sigma$ is the static conductivity and $\epsilon_{P}^{'}+\jmath\epsilon_{P}^{''}(\omega)$
is the complex permittivity resulting from the polarization. As an
illustration, when the material follows the Debye law as for the standard
FR4 epoxy substrate~\citet{Koledisnteva-2002}, one obtains the plots
presented in figure~\ref{fig:DebyeModelFR4} for the real and imaginary
parts of the permittivity.

In this figure one can see that the imaginary part effectively measured
at 1 GHz for instance is nearly thrice the contribution of polarization
because of the conduction. Nevertheless concerning FR4, as the resonance
occurs at high frequency, whereas the conduction contribution decreases
with the frequency, isolating both effect is not really difficult.
It would not be the case if the resonance occurred at 100~MHz for
instance.

The Debye model presented above applies when the considered material
presents only one relaxation. In sophisticated material such as polymers
or cements, empirically modified versions of the Debye law are often
used such as the Cole-Cole model \citet{Cole1941} to describe properly
the superposition of many relaxations at low frequencies. In all cases
the measured permittivity can be written as in equation~(\ref{eq:FullDebye}).

Considering only one relaxation frequency, the parameters of the Debye
model can be estimated through optimization \citet{LJRJ+98}. Considering
equation~(\ref{eq:FullDebye}), one can search for the parameter
set $(p_{1},p_{2},p_{3},p_{4})$ that minimizes the criterion
\begin{equation}
{\cal C}=\sum_{\omega_{i}}\left|\epsilon_{M}(\omega_{i})-\Big(p_{1}+\frac{p_{2}-p_{1}}{1+j\omega_{i}/p_{3}}-\jmath\frac{p_{4}}{\omega_{i}}\Big)\right|^{2}.\label{eq:Criterion}
\end{equation}

In this equation, $\epsilon_{M}(\omega_{i})$ are the measured values
of the permittivity at the circular frequencies $\omega_{i}$. The
minimization of $\mathcal{C}$ gives good results when testing materials
such as FR4 since the relaxation frequency is sufficiently high in
these materials so that its effect can be easily separated from conduction
effect. However it is much less efficient with multiple superimposed
relaxations when the Cole-Cole or more complicated laws \citet{Havriliak1967}
apply.

\section{Estimation of conduction from Kramers-Kronig relations}

The displacement field $\vec{D}$ is the material polarization response
to the electrical field $\vec{E}$. One has
\begin{equation}
\vec{D}_{P}(\omega)=\epsilon_{P}(\omega)\vec{E}(\omega).\label{eq:}
\end{equation}

The inverse Fourier transform of this relation leads to
\begin{equation}
\vec{D}_{P}(t)=\epsilon_{P}(t)*\vec{E}(t)\label{eq:DfromEConvForm}
\end{equation}
where operator $*$ stands for the convolution product.

\subsection{General principle }

The material studied is considered linear and invariant in time. Because
equation~(\ref{eq:DfromEConvForm}) is valid for any temporal evolution
of $\vec{E}(t)$, $\epsilon_{P}(t)$ can be interpreted as an impulse
response. Therefore the causality principle implies:
\begin{equation}
\epsilon_{P}(t)=u(t)\times\epsilon_{P}(t)\label{eq:Causality}
\end{equation}
where $u(t)$ refers to as the unity step function. Using the Fourier
transform and considering the Principal Value, one obtains from (\ref{eq:Causality})
\begin{equation}
\epsilon_{p}(\omega)=-\frac{2\jmath}{\omega}*\epsilon_{P}(\omega)=\jmath\times\mathrm{H_{T}}(\epsilon_{P}(\omega))\label{eq:KKConvolutionForm}
\end{equation}
where $\mathrm{H_{T}}$ is the Hilbert transform. The minus sign in
(\ref{eq:KKConvolutionForm}) may change accordingly to the definition
used for the Fourier transform. In this paper we use the sign that
is relevant with the FFT algorithm. Relation~(\ref{eq:KKConvolutionForm})
is known as the Kramers-Kronig (K-K) relation. It is often referred
to as the K-K relation pair by using separately the real part $\epsilon_{P}'$
and the imaginary part $\epsilon_{P}''$ of $\epsilon_{P}(\omega)$
as
\begin{equation}
\begin{cases}
\epsilon_{P}''(\omega) & ={\rm H_{T}}(\epsilon_{P}'(\omega))\\
\epsilon_{P}'(\omega) & =-{\rm H_{T}}(\epsilon_{P}''(\omega))
\end{cases}\label{eq:Classical_KK}
\end{equation}

Since Equation~(\ref{eq:FullDebye}) can be rewritten as 
\begin{equation}
\epsilon_{M}(\omega)=\epsilon_{\infty}+\epsilon'_{P}(\omega)+\jmath\big(\epsilon_{P}''(\omega)-\frac{\sigma}{\omega}\big)\label{eq:-2}
\end{equation}
and because $\mathrm{H_{T}}(\epsilon_{\infty})=0$ since $\epsilon_{\infty}$
is constant, one has
\begin{equation}
\epsilon_{P}''(\omega)={\rm H_{T}}\left(\Re(\epsilon_{M}(\omega))\right).\label{eq:-3}
\end{equation}
It is thus now possible to retrieve the conductivity $\sigma$ from
the measurements by 
\begin{equation}
\sigma=\jmath\omega\left[\Im((\epsilon_{M}(\omega))-{\rm H_{T}}\left(\Re(\epsilon_{M}(\omega))\right)\right]\label{eq:Conductivity}
\end{equation}

It is possible to verify the consistency of calculation (\ref{eq:Conductivity})
by verifying whether if $\sigma$ is constant over a large range of
frequencies.

\subsection{Calculation using the Fast Fourier Transform}

Hilbert transform is a convolution product. It can therefore be calculated
using a Fourier transform. Numerically speaking, the discrete Hilbert
transform \foreignlanguage{english}{${\rm HT({\rm S})}$} can be computed
for a given sampled signal $S(x_{i})$ using the FFT algorithm as
\begin{equation}
{\rm HT({\rm S})}=IFFT(\jmath\times SGN(k_{i})\times FFT(S))\label{eq:Th_Computation}
\end{equation}

In this equation ${\rm SGN}(k_{i})$ emulates the signum function.
It is $1$ when $k_{i}>0$, $-1$ when $k_{i}<0$ and $0$ when $k_{i}=0$.
It is also possible to use a direct convolution algorithm to compute
the Hilbert transform. It does not yield better results than using
the FFT algorithm and it is much slower.

\section{Application to real measurement\label{sec:Application}}

\begin{figure}[h]
\centering{}\fontsize{8pt}{10pt}\selectfont
\psfrag{Real}{$\Re \big( \epsilon _M (\omega)/\epsilon _0 \big)$}
\psfrag{Imag}{$ \Im \big( \epsilon _M (\omega)/\epsilon _0 \big)$}
\psfrag{F/m}[cc][cc]{}
\psfrag{00}{}
\psfrag{0}{$1$}
\psfrag{1}{$10$}
\psfrag{2}{$10^2$}
\psfrag{3}{$10^3$}
\psfrag{4}{$10^5$}
\psfrag{5}{$10^5$}
\psfrag{6}{$10^6$}
\psfrag{Hz}{Hz}
\psfrag{4.35}{}
\psfrag{3.3}[cc][cc]{}
\psfrag{Polarisation}{Polarization : $\epsilon ' _P (\omega )/\epsilon _0$}
\psfrag{Conduction}{Conduction : $\Im \big( \epsilon _M (\omega)/\epsilon _0 \big)- \epsilon ' _P (\omega )/\epsilon _0$}
\psfrag{EpsilonRelatif}[][]{}
\psfrag{fc=2.63Hz}[][]{fc $\approx$ 2Hz}
\psfrag{1/S=943}[l][l]{Resistivity : 193 $\Omega$m}
\psfrag{S=1.1e-09}[l][l]{}
\includegraphics[width=8cm]{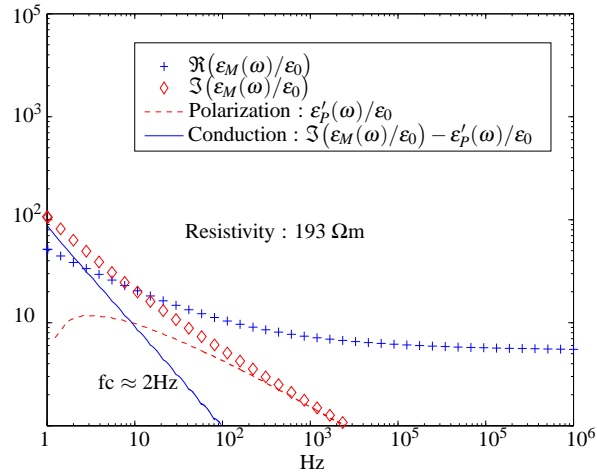}\caption{Measured permittivity on doped cement sample.\label{fig:PermittivityConcrete}}
\end{figure}

We have applied the method presented above to measurements performed
on one-month dry doped cement samples made of MIPLACOL EN 13813-CT16F4
from Bostik SA. The measurements were carried out using a SOLARTRON
1260 impedance meter in the frequency range from 1~Hz up to 1~MHz.
To reduce the measurement time, they are performed at logarithmically
spaced frequencies. In order to use the FFT algorithm, they were evenly
re-sampled at 0.2~Hz using a linear interpolation between measurement
points. The imaginary and real parts of the measured permittivity
are shown in figure~\ref{fig:PermittivityConcrete}.

In this figure, $\epsilon_{0}$ is the vacuum permittivity. Considering
the shape of the real part of the measurement, one can see that the
relaxation circular frequency is very low and therefore superimposes
with the conductivity contribution. This measurement is thus a typical
case where the proposed method presented in this paper is interesting.

The result of the Hilbert transform applied to the real part of the
re-sampled measurements is presented in red dashed line in figure~\ref{fig:PermittivityConcrete}.
It presents a bell like shape as expected. By localizing its maximal
value one can see that relaxations occur around $f_{\mathrm{rel}}=2$~Hz
which is indeed low. The conduction contribution to the imaginary
part of the permittivity is presented in solid blue in the same figure.
A straight line is obtained as expected. One can notice that the contribution
of conduction is not negligible in regard to the overall imaginary
part of the permittivity below 2~Hz.

\begin{figure}[h]
\centering{}\fontsize{8pt}{10pt}\selectfont
\psfrag{-6}{$10^{-4}$}
\psfrag{-8}{$10^{-6}$}
\psfrag{-10}{$10^{-8}$}
\psfrag{-12}{$10^{-10}$}
\psfrag{-14}{$10^{-12}$}
\psfrag{-16}{$10^{-14}$}
\psfrag{0}{$1$}
\psfrag{1}{$10$}
\psfrag{2}{$10^2$}
\psfrag{3}{$10^3$}
\psfrag{4}{$10^5$}
\psfrag{5}{$10^5$}
\psfrag{6}{$10^6$}
\psfrag{Hz}{Hz}
\psfrag{S/m}[][]{S/m}
\psfrag{fc=2.63Hz}[][]{fc=2.63Hz}
\psfrag{sigma}[][]{$\sigma = (5.2 \pm 0.3)\times 10^{-9}$~S/m}
\psfrag{Calcul}[][]{$\sigma$ estimation data}
\includegraphics[width=8cm]{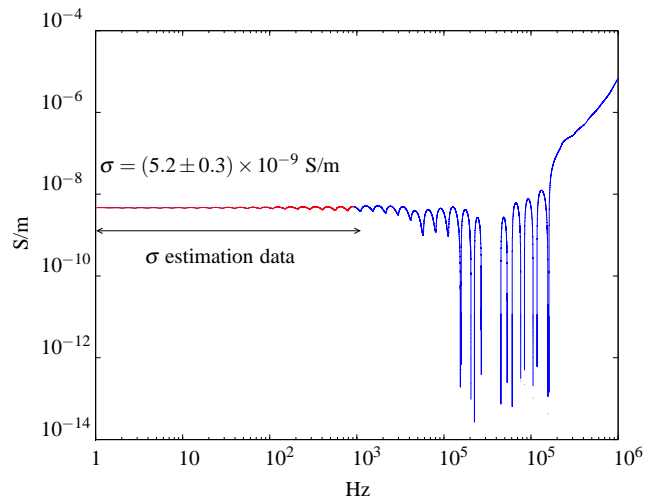}\caption{ Estimated conductivity.\label{fig:ConductivityConcrete}}
\end{figure}

The conductivity calculated using equation~(\ref{eq:Conductivity})
is presented in figure~\ref{fig:ConductivityConcrete}. The values
obtained are roughly the same up to 1~kHz (red part). As the contribution
of conductivity is very small for higher frequencies, the values obtained
above 1~kHz are no longer relevant. At even higher frequency (above
100~kHz) one only obtains the contribution of the multiplication
by $\omega$ of equation~(\ref{eq:Conductivity}). The conductivity
estimated in figure~\ref{fig:ConductivityConcrete} is the mean value
of the curve tagged as $\sigma$ estimation value The standard deviation
to this mean value is also indicated.

It is worth noting that the slope of the asymptotic behavior at infinity
of the polarization of figure~\ref{fig:PermittivityConcrete} (dashed
red line) is less than 20~dB/decade. This means that the Debye model
does not strictly apply.

\section{Conclusion}

In this paper, the nature of the permittivity measurement is discussed
to show that it can always be split between conduction and polarization
effects. Because of the causal aspect of polarization, which is the
material response to the electrical field, we have proposed a method
to extract the material conductivity independently from any behavioral
models of that material. This makes the method interesting compared
to optimization or fitting methods which require a model. The method
is applied to complex cement samples and the results show its efficiency
to retrieve the value of the conductivity of such samples. In MIPLACOL
EN 13813-CT16F4, the conductivity estimated to be $5.2\times10^{-9}$~S/m

\section*{References}

\bibliographystyle{unsrt}

\begin{thebibliography}{10}

\bibitem{Agilent2013}
{\em Agilent {I}mpedance {M}easurement {H}andbook}.
\newblock Agilent Technologies, fourth edition, October 2013.
\newblock www.agilent.com/find/impedance.

\bibitem{Kittel1976}
C.~Kittel.
\newblock {\em Introduction to solid state physics}.
\newblock John Willey \& sons, fifth edition, 1976.

\bibitem{Jackson1967}
{Jackson J.D}.
\newblock {\em Classical {E}lectrodynamics}.
\newblock John Wiley \& sons, sixth edition, 1967.
\newblock Library of Congress Catalog Card Number: 62-8774.

\bibitem{Ramo1984}
{Ramo S}, {Whinnery J.R}, and {Van Duzer T}.
\newblock {\em Fields and waves in communication electronics}.
\newblock John Wiley and Sons, 1984.

\bibitem{BOHREN2010}
{C.F. Bohren}.
\newblock What did {K}ramers and {K}ronig do and how did they do it ?
\newblock {\em Eur. J. Phys}, 31:573--577, 2010.
\newblock doi : 1088/0143-0807/31/3/014.

\bibitem{Lucas2012}
{J. Lucas}, {E. G\'eron}, {T. Ditchi}, and {S. Hol\'e}.
\newblock A fast {F}ourier transform implementation of the {K}ramers-{K}ronig
  relations: {A}pplication to anomalous and left handed propagation.
\newblock {\em AIP Advances}, 2012.
\newblock AIP Advances 2, 032144 (2012); doi: 10.1063/1.4747813.

\bibitem{Koledisnteva-2002}
{M.Y Kolendisteva}, {K.N Rosanov}, {A. Orlandi}, and {J.L Droniak}.
\newblock Extraction of {L}orentzian and {D}ebye of dielectric and magnetic
  dispersive materilas for {FDTD} modeling.
\newblock {\em Journal of Electrical Engineering}, 53(9/S):97--100, 2002.

\bibitem{Cole1941}
{S. K. Cole} and {H. R. Cole}.
\newblock Dispersion and {A}bsorption in dielectrics.
\newblock {\em Journal of Chemical physics}, 9:341--351, April 1941.

\bibitem{LJRJ+98}
{Lagarias J.C}, {Reeds J.A}, {Wright M.H}, and {Wright P.E}.
\newblock Convergence {P}roperties of the {N}elder-{M}ead {S}implex {M}ethod in
  {L}ow {D}imensions.
\newblock {\em SIAM Journal of Optimization}, 9(1):112--147, 1998.

\bibitem{Havriliak1967}
{S. Hawriliak} and S.Negami.
\newblock A complex plane representation of dielectric and mechanical
  relaxation processes in some polymers.
\newblock {\em Polymer}, 8:161--210, 1967.

\end{thebibliography}

\end{document}